\documentclass[a4paper,11pt]{article}

\pdfoutput=1 

\usepackage{jcappub} 

\usepackage[T1]{fontenc} 

\usepackage{graphicx} 
\usepackage{amsmath}
\usepackage{amssymb}
\usepackage[utf8]{inputenc}

\usepackage{lineno}

\usepackage[normalem]{ulem}
\usepackage{xcolor}

\title{Solar axion searches with RES-NOVA: projected sensitivity and first prototype limit}

\author[a,b]{D.~Alloni,}
\author[c,d]{G.~Benato,}
\author[e,f]{P.~Carniti,}
\author[e,f]{M.~Cataldo,}
\author[g]{L.~Chen,}
\author[e,f]{M.~Clemenza,}
\author[e,f]{M.~Consonni,}
\author[e,f]{G.~Croci,}
\author[h]{I.~Dafinei,}
\author[i,j]{F.A.~Danevich,}
\author[k,l,m]{J.~De~Miguel,}
\author[b]{C.~de~Vecchi,}
\author[e,f]{D.~Di~Martino,}
\author[d,n]{R.~Elleboro,}
\author[e,f]{N.~Ferreiro~Iachellini,}
\emailAdd{nahuel.ferreiroiachellini@unimib.it}
\author[c,h]{F.~Ferroni,}
\author[f,o]{F.~Filippini,}
\author[p]{V.~Fonoll,}
\author[q]{S.~Ghislandi,}
\author[e,f]{A.~Giachero,}
\author[p,r]{M.~Giannotti,}
\emailAdd{mgiannotti@unizar.es}
\author[e,f]{L.~Gironi,}
\author[d]{P.~Gorla,}
\author[e,f]{C.~Gotti,}
\author[d]{D.L.~Helis,}
\author[i]{D.V.~Kasperovych,}
\author[i]{V.V.~Kobychev,}
\author[u,v]{A.~Lella,}
\emailAdd{alessandro.lella@unipd.it}
\author[w]{G.~Lucente,}
\emailAdd{lucenteg@slac.stanford.edu}
\author[e,f]{G.~Marcucci,}
\author[d,n]{A.~Melchiorre,}
\author[s,b]{A.~Menegolli,}
\author[d]{S.~Nisi,}
\author[t]{M.~Musa,}
\author[c,d]{L.~Pagnanini,}
\author[e,f]{L.~Pattavina,}
\author[f]{G.~Pessina,}
\author[d]{S.~Pirro,}
\author[e,f]{S.~Pozzi,}
\author[b]{M.C.~Prata,}
\author[d]{A.~Puiu,}
\author[e,f]{S.~Quitadamo,}
\author[t]{M.P.~Riccardi,}
\author[b]{M.~Rossella,}
\author[s,b]{R.~Rossini,}
\author[x,f]{E.~Sala,}
\author[f,o]{F.~Saliu,}
\author[a]{A.~Salvini,}
\author[i,j]{V.I.~Tretyak,}
\author[e,f]{L.~Trombetta,}
\emailAdd{l.trombetta7@campus.unimib.it}
\author[e,f]{D.~Trotta,}
\author[g]{H.~Yuan}

\affiliation[a]{Laboratorio Energia Nucleare Applicata, Via Aselli 41, I-27100 Pavia, Italy}
\affiliation[b]{INFN Sezione di Pavia, Via Bassi 6, I-27100 Pavia, Italy}
\affiliation[c]{Gran Sasso Science Institute, Viale F. Crispi 7, I-67100 L'Aquila, Italy}
\affiliation[d]{INFN Laboratori Nazionali del Gran Sasso, Via G. Acitelli 22, I-67100 Assergi, Italy}
\affiliation[e]{Dipartimento di Fisica, Universit\`a di Milano - Bicocca, Piazza della Scienza 3, I-20126 Milano, Italy}
\affiliation[f]{INFN Sezione di Milano - Bicocca, Piazza della Scienza 3, I-20126 Milano, Italy}
\affiliation[g]{Shanghai Institute of Ceramics, CAS, 1295 Dingxi Road, Shanghai 200050, P.R. China}
\affiliation[h]{INFN Sezione di Roma, P.le Aldo Moro 2, I-00185 Roma, Italy}
\affiliation[i]{Institute for Nuclear Research of NASU, 03028 Kyiv, Ukraine}
\affiliation[j]{Institute of Experimental and Applied Physics, Czech Technical University in Prague, 11000 Prague, Czech Republic}
\affiliation[k]{Instituto de Astrof\'isica de Canarias, E-38200 La Laguna, Tenerife, Spain}
\affiliation[l]{Departamento de Astrof\'isica, Universidad de La Laguna, E-38206 La Laguna, Tenerife, Spain}
\affiliation[m]{The Institute of Physical and Chemical Research (RIKEN), Center for Advanced Photonics, 519-1399 Aramaki-Aoba, Aoba-ku, Sendai, Miyagi 980-0845, Japan}
\affiliation[n]{Dipartimento di Scienze Fisiche e Chimiche, Universit\`a degli Studi dell'Aquila, I-67100 L'Aquila, Italy}
\affiliation[o]{DISAT, Universit\`a di Milano - Bicocca, Piazza della Scienza 1, I-20126 Milano, Italy}
\affiliation[p]{Centro de Astrof\'{i}sica y F\'{i}sica de Altas Energ\'{i}as (CAPA), University of Zaragoza, Zaragoza 50009, Arag\'{o}n, Spain}
\affiliation[q]{Massachusetts Institute of Technology, Cambridge, MA 02139, USA}
\affiliation[r]{Department of Chemistry and Physics, Barry University, Miami Shores, FL 33161, United States of America}
\affiliation[s]{Dipartimento di Fisica, Universit\`a di Pavia, Via Bassi 6, I-27100 Pavia, Italy}
\affiliation[t]{Dipartimento di Scienze della Terra e dell'Ambiente, Universit\`a di Pavia, Via Ferrata 7, I-27100 Pavia, Italy}
\affiliation[u]{Dipartimento di Fisica e Astronomia, Universit\`a degli Studi di Padova, Via Marzolo 8, 35131 Padova, Italy}
\affiliation[v]{Istituto Nazionale di Fisica Nucleare (INFN), Sezione di Padova, Via Marzolo 8, 35131 Padova, Italy}
\affiliation[w]{SLAC National Accelerator Laboratory, Stanford University, Menlo Park, CA 94025}
\affiliation[x]{Center for Underground Physics, Institute for Basic Science, 34126 Daejeon, Korea}

\abstract{RES-NOVA is a cryogenic experiment based on PbWO$_4$ scintillating bolometers produced from archaeological lead, originally designed to detect coherent elastic neutrino-nucleus scattering (CE$\nu$NS) from galactic core-collapse supernovae. We show that the same high-density, high-$Z$ absorbers and sub-keV energy resolution make RES-NOVA an appealing probe of solar axions. We compute the expected signal from the Primakoff, ABC (atomic recombination and de-excitation, Bremsstrahlung, Compton), longitudinal-plasmon (LP), and $^{57}$Fe nuclear-line components of the solar axion flux, folded through the inverse-Primakoff and axioelectric detection channels in PbWO$_4$, which probe the axion couplings to photons ($g_{a\gamma}$), electrons ($g_{ae}$), and nucleons ($g_{aN}$), respectively. We derive projected sensitivities for a $1\,\mathrm{ton\cdot y}$ exposure of the RES-NOVA demonstrator in the $(g_{ae},g_{a\gamma})$, $(g_{ae},g_{aN})$, and $(g_{a\gamma},g_{aN})$ planes. The projected reach in $g_{ae}$ approaches that of XENONnT to within a factor of $\sim2$, despite a background roughly four orders of magnitude higher, thanks to the large target mass, the high-$Z$ enhancement of the axioelectric and inverse-Primakoff cross sections, and the sub-keV energy resolution. Because the inverse-Primakoff constraint on $g_{a\gamma}$ relies on absorption rather than coherent conversion, it is independent of the axion mass, unlike bounds from magnetic helioscopes. We complement these projections with the first solar-axion exclusion limit obtained on real data with a 13~g PbWO$_4$ prototype grown from archaeological lead, using a background-model-independent optimum-interval analysis. 
These results establish \mbox{RES--NOVA} as a promising multi-coupling probe of solar axions, with simultaneous sensitivity to $g_{a e}, g_{a \gamma}$ and $g_{a N}$ that is complementary to existing helioscope and direct-detection searches. }
\begin{document}
\maketitle
\flushbottom

\section{Introduction}
\label{sec:intro}

The strong CP problem---the striking absence of observable CP violation in the
strong interactions, despite the presence of a CP-odd $\theta$-term allowed in the
QCD Lagrangian---remains one of the outstanding puzzles of the Standard Model. Its
most compelling solution is the Peccei--Quinn (PQ) mechanism~\cite{Peccei:1977hh},
which promotes $\theta$ to a dynamical field and predicts a light
pseudo-Nambu--Goldstone boson, the QCD axion~\cite{Weinberg:1977ma,Wilczek:1977pj}. A
defining feature of the QCD axion is that both its mass and its couplings to Standard
Model particles are controlled by a single scale, the PQ symmetry-breaking scale
$f_a$, so that its phenomenology is governed by essentially one parameter. More
general axion-like particles (ALPs), which arise generically in extensions of the
Standard Model and in string compactifications, relax this strict mass--coupling
relation and populate a much broader region of parameter
space~\cite{DiLuzio:2020wdo,Cicoli:2026fqp}. Beyond their original motivation, axions
and ALPs have emerged as leading candidates for cold dark matter (CDM) and as possible
mediators of new feeble interactions, making their search one of the most active
frontiers at the interface of particle physics, astrophysics and
cosmology~\cite{Adams:2022pbo,Giannotti:2024xhx,Arza:2026rsl}.

The specific values of the axion couplings depend on the ultraviolet completion of the
PQ sector. Two benchmark realizations are widely adopted: the hadronic KSVZ
model~\cite{Kim:1979if,Shifman:1979if}, in which the axion couples to photons and
nucleons but acquires a coupling to electrons only at the loop level, and the DFSZ
model~\cite{Zhitnitsky:1980tq,Dine:1981rt}, in which the axion couples to charged
leptons already at tree level. As a consequence, the axion--electron coupling
$g_{ae}$ is strongly model dependent: it is naturally suppressed in hadronic
(KSVZ-like) scenarios and sizeable in non-hadronic (DFSZ-like) ones, whereas the
couplings to photons and nucleons are present in both classes. An experiment sensitive
to several couplings simultaneously can therefore, in principle, discriminate between
these classes of models~\cite{DiLuzio:2020wdo,Giannotti:2024xhx,Carenza:2024ehj}.

Among the possible astrophysical sources, the Sun is an exceptionally bright and
well-calibrated axion factory. Solar axions are produced in the high-temperature
plasma of the solar interior through several mechanisms, governed by the axion
couplings to photons ($g_{a\gamma}$), electrons ($g_{ae}$) and nucleons
($g_{aN}$)~\cite{DiLuzio:2021qct,Carenza:2024ehj}. 
These processes yield a quasi-continuous spectrum in
the keV range, with characteristic energies set by the solar core temperature,
together with line-like contributions from specific atomic and nuclear transitions.
Because the resulting flux is essentially independent of the cosmological abundance of
axions, solar searches constrain the axion couplings directly and provide information
complementary to that of haloscopes, which instead assume that axions make up the local
dark matter~\cite{Adams:2022pbo}.

Traditionally, solar-axion searches have been dominated by magnetic helioscopes such as
CAST~\cite{CAST:2024} and the upcoming IAXO~\cite{IAXO:2025ltd}, which convert solar
axions into detectable X-rays in a strong laboratory magnetic field. This coherent
conversion is highly effective but efficient only for sufficiently light axions,
$m_a \lesssim 0.02$~eV, above which the loss of coherence degrades the sensitivity
except in dedicated buffer-gas configurations. A complementary strategy exploits the
direct absorption of solar axions in massive targets: the axioelectric
effect~\cite{Derevianko:2010kz} for the axion--electron coupling, and the inverse
Primakoff conversion in the atomic electric field of the target nuclei~\cite{Abe:2020mcs}
for the axion--photon coupling. Unlike helioscope conversion, these absorption
processes require no macroscopic coherence baseline and are therefore insensitive to
the axion mass over the entire range of solar production. This approach underlies the
solar-axion limits set by large liquid-xenon detectors such as XENONnT~\cite{XENON:2022ltv},
whose ultra-low backgrounds currently provide the most stringent direct bounds on
$g_{ae}$. Their analysis thresholds, however, lie around $1$~keV, leaving the deep
sub-keV region---where part of the solar-axion spectrum resides---essentially
unexplored. Cryogenic calorimeters, which can reach thresholds of
$\mathcal{O}(10\text{--}100)$~eV, are naturally suited to access this window.

In this context, RES-NOVA represents a novel platform for solar-axion detection.
Originally conceived to observe coherent elastic neutrino--nucleus scattering
(CE$\nu$NS) from galactic core-collapse supernovae~\cite{Pattavina:2020cqc}, RES-NOVA is
based on archaeological-lead PbWO$_4$ scintillating bolometers read out by Transition
Edge Sensors (TESs) operated at millikelvin temperatures. The same features that make
this technology attractive for neutrino physics---high-density, high-$Z$ absorbers, the
intrinsically low radioactivity of archaeological lead~\cite{RES-NOVACollaboration:2025stq},
and an excellent energy resolution combined with a low energy threshold---also make it a
compelling probe of solar axions. The high nuclear charge of Pb and W strongly enhances
both the axioelectric and the inverse-Primakoff absorption cross sections, while the
sub-keV resolution allows the distinctive spectral features of the different solar-axion
components---the Primakoff and ABC continua, the forest of atomic recombination lines,
the $^{57}$Fe nuclear line at $14.4$~keV, and the longitudinal-plasmon (LP)
peak---to be resolved on top of a non-flat radiogenic background. Crucially, because the
same crystal is simultaneously sensitive to $g_{a\gamma}$ and $g_{ae}$,
RES-NOVA can constrain several couplings at once and thereby help discriminate between
hadronic and non-hadronic axion models. The demonstrated feasibility of the technology,
with $\mathcal{O}(100\,\mathrm{eV})$ thresholds already achieved on small
prototypes~\cite{FerreiroIachellini:2021qgu}, further opens the possibility of a
sub-$200$~eV threshold that would expose the LP component and, with it, a unique
observational handle on the magnetic-field structure of the deep solar interior ~\cite{OHare:2020wum}---a
regime hardly accessible to standard helioseismology.

In this work we present a comprehensive study of the sensitivity of the RES-NOVA
demonstrator to solar axions. We compute the expected signal from the Primakoff, ABC,
LP and $^{57}$Fe components, folded through the inverse-Primakoff and axioelectric
detection channels in PbWO$_4$, and evaluate the experimental reach with a
profile-likelihood-ratio analysis on an Asimov dataset in the
$(g_{ae},\,g_{a\gamma})$, $(g_{ae},\,g_{aN}^{\rm eff})$ and $(g_{a\gamma},\,g_{aN}^{\rm eff})$ coupling
planes. The axion--nucleon interaction enters only through the production of the $^{57}$Fe $14.4$~keV line, which is governed by a specific effective combination of the proton and neutron couplings, $g_{aN}^{\rm eff}$---dominated by the coupling to neutrons---defined in Sec.~\ref{subsec:Fe57}. We further report the first solar-axion exclusion limit obtained on real data
with a $13$~g PbWO$_4$ prototype grown from archaeological
lead~\cite{RES-NOVA:2026prototype}, derived with a
background-model-independent optimum-interval analysis. Together, these results
position RES-NOVA---en route to the full $1.8$~ton experiment---as a competitive,
multi-coupling solar-axion probe.

\section{Solar Axion Fluxes}

The solar axion flux, $\Phi_a$, receives contributions from several production channels in the solar interior. The dominant production channels from the stellar plasma give rise to a quasi-thermal spectrum, with characteristic energies set by the solar core temperature, $T_c \sim {\rm keV}$. In addition, the spectrum may contain subdominant but potentially relevant non-thermal or line-like contributions associated with specific atomic and nuclear transitions and reactions. The relative importance of these components depends on the axion couplings to photons, electrons, and nucleons, namely $g_{a\gamma}$, $g_{ae}$, and $g_{aN}$, with $N=p,\,n$ for protons and neutrons, respectively. 

In this work, we consider axions with mass $m_a \lesssim T_c$, estimating their emission spectrum with the updated Standard Solar Model (SSM) inputs and the prescriptions of Ref.~\cite{Carenza:2024ehj}. The Boltzmann suppression starts at around $3 T_c$. The resulting flux is shown in Fig.~\ref{fig:allFluxes}.

\subsection{Fluxes from Axion-Photon Coupling}
Through their coupling to photons, light axions are produced in the stellar interior via two dominant channels: the Primakoff flux, from the conversion of thermal photons into axions in the Coulomb fields of the plasma, and the longitudinal-plasmon (LP) flux, from the conversion of longitudinal plasmons into axions in the large-scale solar magnetic field~\cite{Guarini:2020hps,Caputo:2020quz}.

\begin{figure}[h!]
\centering
\includegraphics[width=.89\textwidth]{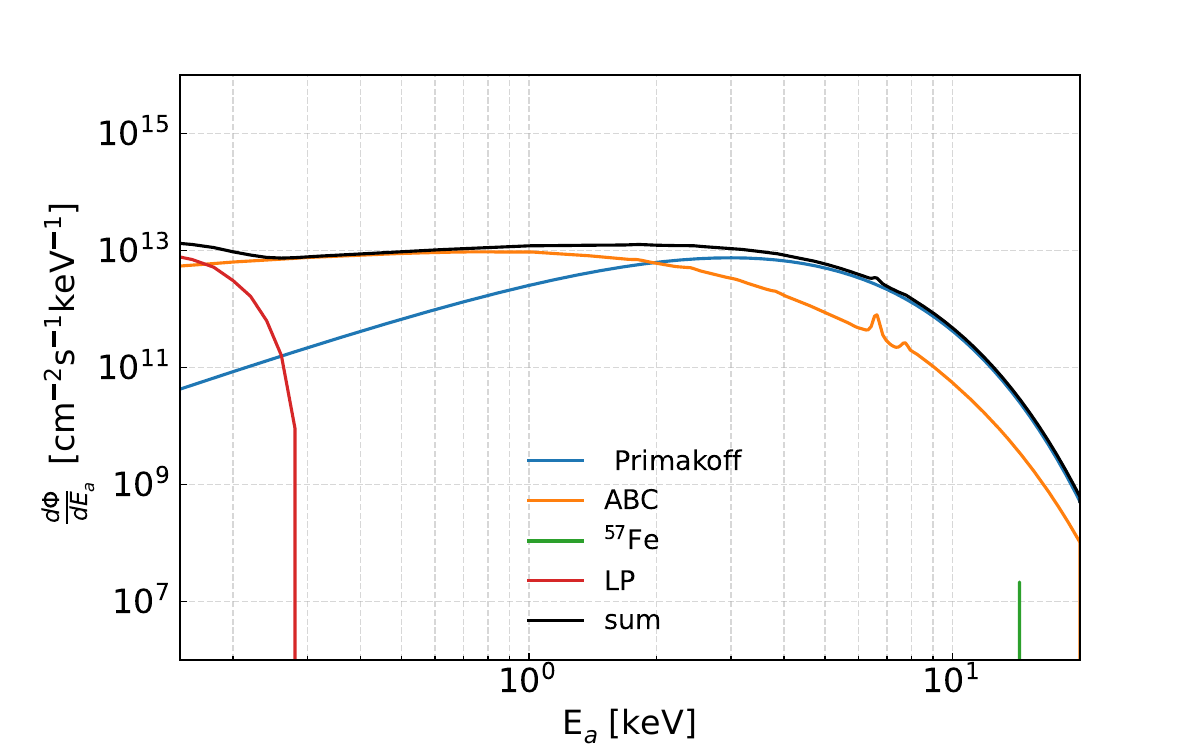}
\caption{Solar axion fluxes at Earth as a function of energy for the production
channels considered in this work: the Primakoff and longitudinal-plasmon (LP) fluxes
governed by $g_{a\gamma}$, the ABC flux governed by $g_{ae}$, and the
monochromatic $^{57}$Fe line at $14.4$~keV governed by $g_{aN}^{\rm eff}$; the black curve is
their sum. Each component is shown for a representative value of the corresponding
coupling. The continuous fluxes peak below $\simeq 5$~keV, while the LP contribution is
confined to the deep sub-keV region ($E \lesssim 200$~eV). The couplings were set to $(g_{a\gamma}, g_{ae}, g_{aN}^{\rm eff}) = (10^{-9}\,{\rm GeV}^{-1}, 10^{-11}, 10^{-9})$.\label{fig:allFluxes}}
\end{figure}

The Primakoff emission produces a quasi-thermal spectrum which we parametrize using a Gamma-distribution function
\begin{equation}
    \frac{d\Phi_P}{dE_a} = C_P \left( \frac{g_{a\gamma}}{10^{-12} \text{ GeV}^{-1}} \right)^2 \left( \frac{E_a}{E_P}\right)^{\beta_P} \exp \left[ {-\left( 1+\beta_P \right) \frac{E_a}{E_P}} \right] \, ,\label{eq:Pflux}
\end{equation}
where $E_a$ is the axion energy in keV, and $C_P$, $E_P$ and $\beta_P$ are free fitting parameters. We fix the values of $C_P$, $E_P$ and $\beta_P$ following the updated SSM parameters reported in Ref.~\cite{Carenza:2024ehj} (see Table~\ref{tab:params}). The resulting flux for $g_{a\gamma}=10^{-9}$\, GeV$^{-1}$ is shown as a blue line in Fig.~\ref{fig:allFluxes}.
The LP flux, by contrast, is resonantly enhanced when the energy of the emitted axion matches the plasma frequency $\omega_{\rm pl}\sim\mathcal{O}(100)\,{\rm eV}$ in the sun, giving rise to a distinctive sub-keV spectral component whose shape is governed by $\omega_{\rm pl}$ and the magnetic-field profile of the solar interior. Given the uncertainty on the latter, we model the LP emission spectrum assuming the average value of the flux reported in Fig.~17 of Ref.~\cite{Carenza:2024ehj}. Since the solar plasma frequency remains below
a few $100$~eV, this contribution is confined to the deep sub-keV region and
is accessible only to detectors with thresholds of order $200$~eV or below. A
statistically significant detection of this component could therefore provide a
unique observational handle on the magnetic-field structure of the deep solar
interior---a regime that remains difficult to probe with standard
helioseismology.

\subsection{Fluxes from Axion-Electron Coupling}
If the axion--electron coupling $g_{ae}$ is non-zero, an additional source of
solar axions is the ABC flux, which collects the contributions from Atomic
recombination and de-excitation (A), Bremsstrahlung (B), and Compton scattering
(C)~\cite{Redondo:2013wwa,Carenza:2024ehj}. The B and C processes proceed through the
interaction of axions with the electrons of the solar plasma and produce a
high-intensity, continuous quasi-thermal flux. The A component instead appears as
a dense set of characteristic atomic emission lines superimposed on the B and C
continuum.

\begin{table}[t]
\centering
\caption{Best-fit parameters~\cite{Carenza:2024ehj} for the different solar axion production channels leading to a quasi-thermal spectrum. The values of $C_i$ are quoted for reference couplings $g_{a\gamma}=10^{-12}$\,GeV$^{-1}$ and $g_{ae}=10^{-12}$.}
\begin{tabular}{lcccc}
\hline
Channel & Coupling & $C_i$ [cm$^{-2}$ s$^{-1}$ keV$^{-1}$] & $E_i$ [keV] & $\beta_i$ \\
\hline
Primakoff ($i=$P) & $g_{a\gamma}$ & $2.2 \times10^{8}$  & $4.2$ & $2.5$ \\
Bremsstrahlung ($i=$B) & $g_{a e}$ & $3.8\times10^{11}$ & $1.6$ & $0.8 $ \\
Compton ($i=$C) & $g_{a e}$ & $8.8 \times10^{11}$     & $5.1$ & $3.0$ \\
\hline
\end{tabular}
\label{tab:params}
\end{table}
The two continuous components have the same spectral shape as the Primakoff flux and can therefore be parametrized by the Gamma-distribution form introduced in Eq.~\eqref{eq:Pflux}. In this case, their normalization is set by the axion-electron coupling, $g_{ae}$, while the best-fit parameters $(C_i, E_i, \beta_i)$ are specific to each production process. Explicitly,
\begin{equation}
    \frac{d\Phi_i}{dE_a} = C_i \left( \frac{g_{ae}}{10^{-12}} \right)^2
    \left( \frac{E_a}{E_i}\right)^{\beta_i}
    \exp\left[ -\left( 1+\beta_i \right) \frac{E_a}{E_i} \right] \, ,
    \qquad i = \mathrm{B},\,\mathrm{C} \, ,
    \label{eq:ABCflux}
\end{equation}
where the coefficients of Table~\ref{tab:params} are quoted for a reference coupling $g_{ae}=10^{-12}$. The
resulting B$+$C flux peaks around $1$~keV, in contrast to the Primakoff peak at $\simeq 4.17$~keV. The A component cannot be captured by a smooth parametrization, as it consists of
a forest of narrow atomic lines. We therefore adopt the numerical computation of Ref.~\cite{Redondo:2013wwa} for this contribution.

\subsection{The $^{57}$Fe Nuclear Line}
\label{subsec:Fe57}
Finally, we account for the monochromatic flux produced by the M1 nuclear transition of $^{57}$Fe at 14.4~keV. This emission is governed by an effective
ALP--nucleon coupling $g_{aN}^{\mathrm{eff}} = 0.16\,g_{ap} + 1.16\,g_{an}$~\cite{DiLuzio:2021qct},
which is predominantly sensitive to the axion--neutron interaction~\cite{Haxton:1991pu,Avignone:2017ylv}. Here we follow Ref.~\cite{DiLuzio:2021qct}, whose updated solar model and nuclear
matrix elements yield a total $^{57}$Fe axion flux at Earth
\begin{equation}
\Phi_a = 5.06\times10^{23}\,\left(g_{aN}^{\mathrm{eff}}\right)^{2}
~\mathrm{cm^{-2}\,s^{-1}} .
\end{equation}
The natural line width is negligible; the spectral shape is set by the thermal
Doppler broadening of the emitting $^{57}$Fe nuclei in the solar core,
\begin{equation}
\sigma(T) = E_0\sqrt{\frac{T}{m_{^{57}\mathrm{Fe}}}} \simeq 2~\mathrm{eV}
\qquad (\mathrm{FWHM}\simeq 5~\mathrm{eV}),
\end{equation}
with $E_0 = 14.4$~keV and $T\sim1.3$~keV the solar-core temperature. We therefore
model the differential flux as a narrow Gaussian centred at $E_0$,
\begin{equation}
\frac{d\Phi_a (E_a)}{dE_a} = \Phi_a\,\frac{1}{\sqrt{2\pi}\,\sigma}
\exp\!\left[-\frac{(E_a-E_0)^2}{2\sigma^{2}}\right] .
\end{equation}
The finite detector energy resolution, much larger than $\sigma$, is applied at a
later stage by convolving the full spectrum with the detector response. 

\section{The RES-NOVA detector}
The RES-NOVA experiment exploits cryogenic calorimetry to achieve the high energy resolution and low thresholds required for rare event searches. The detection principle relies on the use of PbWO$_4$ crystals as absorbers, characterized by an extremely low heat capacity ($C$) when operated at millikelvin temperatures~\cite{kg-scale}. 

The detector is planned to be hosted at the underground Gran Sasso Laboratory of INFN (Italy). The site provides an overburden of approximately 3600~m w.e., offering an exceptional suppression of the cosmic-ray-induced background \cite{G.Bellini_2012}, which is a fundamental requirement for investigating low-rate phenomena like the solar axions emission.

The signal in RES-NOVA is purely thermal. When a particle, such as a solar axion, interacts within the PbWO$_4$ crystal, it deposits an energy $\Delta E$ that results in a measurable temperature increase $\Delta T \propto \Delta E/C$. To precisely reconstruct these temperature variations, RES-NOVA utilizes Transition Edge Sensors (TESs). The feasibility of this technology has been demonstrated using small-scale prototypes ($\sim$15~g), achieving $\mathcal{O}(100~$eV$)$ thresholds~\cite{FerreiroIachellini:2021qgu}.

\begin{figure}[t]
\centering
\includegraphics[width=.89\textwidth]{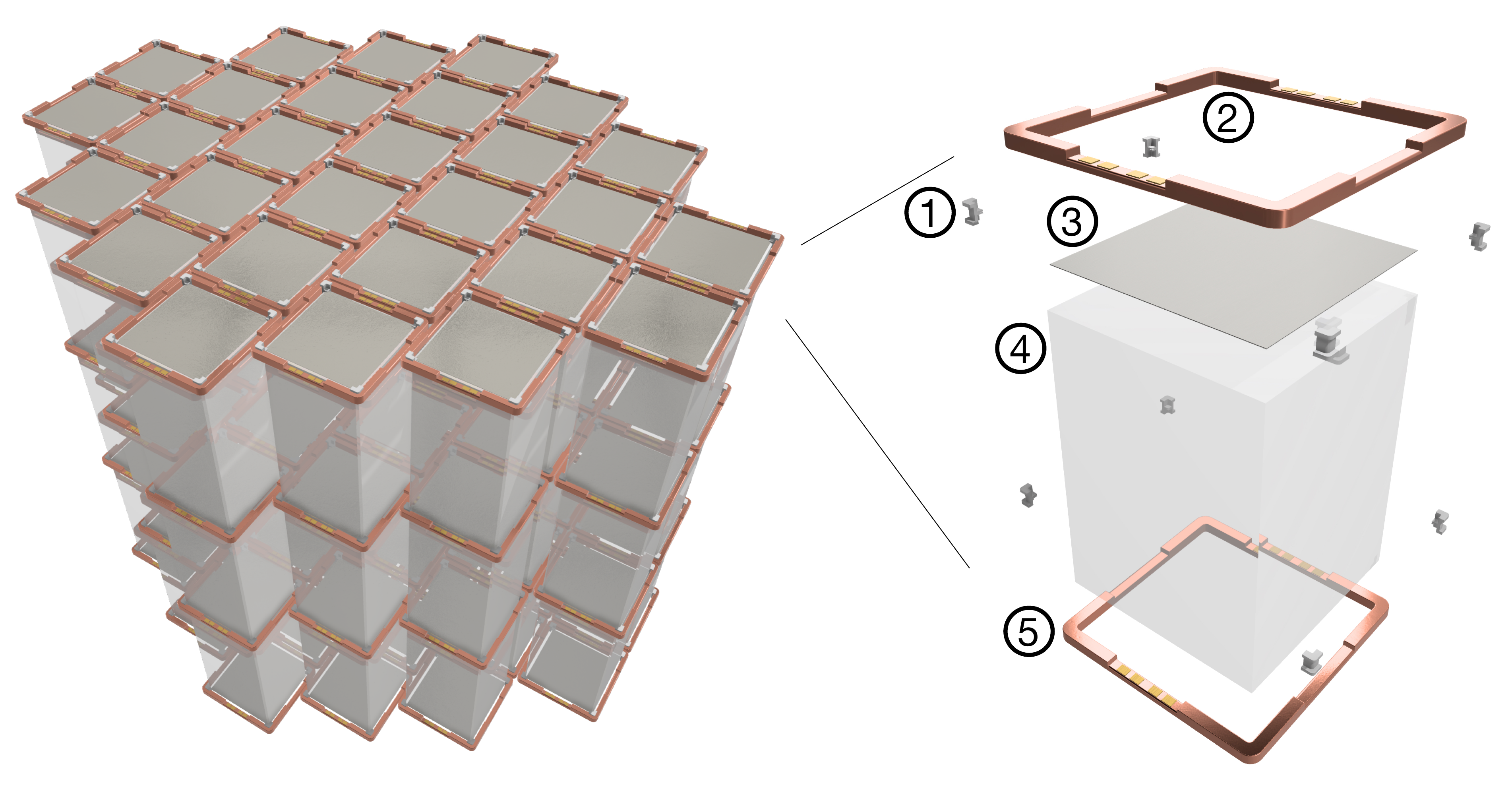}
\caption{Schematic representation of the RES-NOVA demonstrator. \emph{(Left)} The full
array of $84$ PbWO$_4$ modules, for a total active mass of $170$~kg. \emph{(Right)}
Exploded view of a single module, where the numbered elements indicate
(1)~PTFE holding clamps, (2)~Cu top frame, (3)~light detector absorber,
(4)~PbWO$_4$ crystal, and (5)~Cu bottom frame.\label{fig:det}}
\end{figure}

The RES-NOVA demonstrator, shown in Fig.~\ref{fig:det}, will consist of 84 modules for a total active mass of about 170~kg, serving as a functional precursor to the 1.8~t RES-NOVA experiment \cite{Pattavina:2020cqc}. While the module design includes a secondary light detector for particle identification (crucial for nuclear recoil identification, and background rejection~\cite{Beeman:2012wz}), the study of solar axions focuses on the primary thermal signal, as these interactions manifest as electron recoils ($\beta$/$\gamma$-like events). The background level is estimated with a detailed Monte Carlo simulation and its plateau is approximately 1~dru \cite{RES-NOVACollaboration:2025stq} (differential rate unit, 1~dru = 1~count/keV/kg/day).

\section{Solar axion detection channels}
In this work, the sensitivity of RES-NOVA to axions is assessed with two primary detection channels: the inverse Primakoff, induced by the axion-photon coupling; and the axioelectric effect, driven by the axion-electron coupling. The corresponding cross-sections are determined by the same couplings responsible for axion production in the stellar interior, leading to a non-trivial interplay in the expected event rate. 

\subsection{Inverse Primakoff Effect}

Axions coupled to photons can be converted into X-rays in the Coulomb field of atomic nuclei with charge $Ze$ via the inverse Primakoff process,
\begin{equation}
    a + Ze \rightarrow \gamma + Ze \, .
\end{equation}
This channel provides an efficient detection mechanism in high-$Z$ materials such as PbWO$_4$, as the large nuclear charge enhances the interaction probability.

In the limit of relativistic axions ($E_a \gg m_a$), the differential cross section can be written as~\cite{Abe:2020mcs} 
\begin{equation}
    \frac{d\sigma_{P}}{d\Omega} = \frac{\alpha g_{a\gamma}^2}{16\pi}
    \frac{\sin^2\theta}{(1-\cos\theta)^2}
    \left| Z - F(q) \right|^2 \, ,
    \label{eq:invP_dsigma}
\end{equation}
where $\theta$ is the scattering angle and $\alpha$ is the fine-structure constant. The momentum transfer is given by
\begin{equation}
    q = 2 E_a \sin\left(\frac{\theta}{2}\right) \, ,
\end{equation}
while the factor $|Z - F(q)|^2$ accounts for the screening of the nuclear charge by atomic electrons. At low momentum transfer, corresponding to forward scattering ($\theta \to 0$), the atomic electrons screen the nuclear charge and suppress the interaction, while at large momentum transfer the full nuclear charge $Z$ is recovered. This behavior regulates the otherwise divergent forward scattering contribution arising from the $(1-\cos\theta)^{-2}$ term.

\begin{figure*}[]
\centering
\includegraphics[width=.44\textwidth]{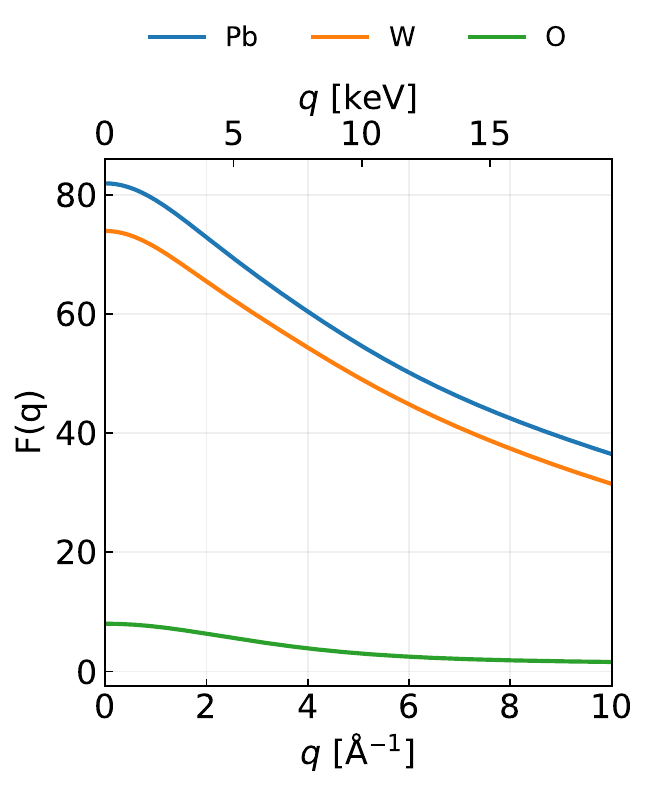} \;\hfill
\includegraphics[width=.49\textwidth]
{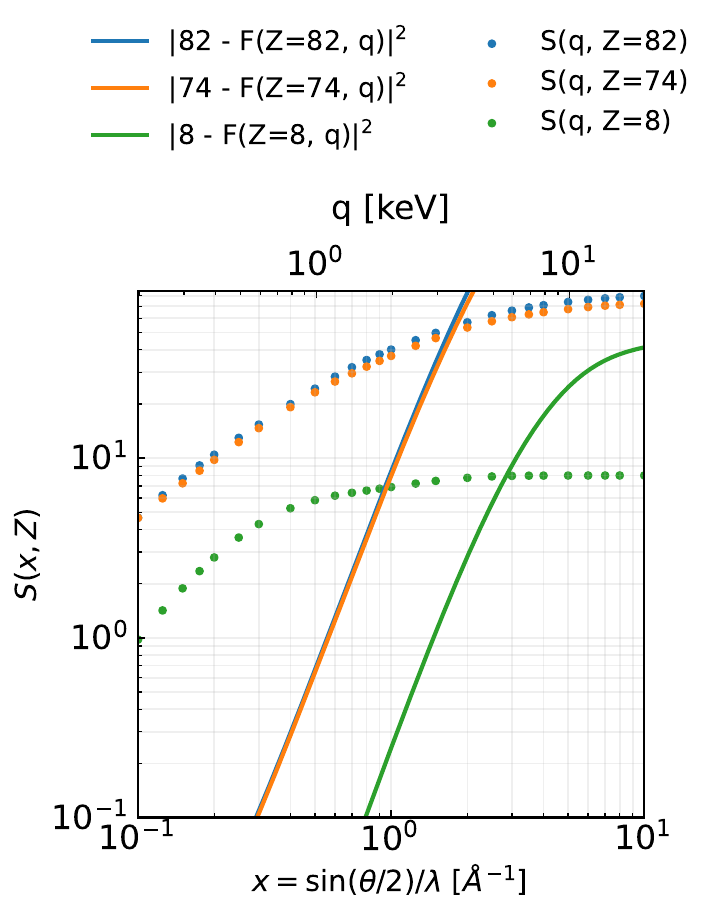}
\caption{\textit{(Left)} Atomic form factors for elements in PbWO$_4$. The atomic form factors $F(q, Z)$ for Lead, Tungsten, and Oxygen were calculated using the relativistic Hartree-Fock coefficients provided by the International Tables for Crystallography~\cite{ITC_VolC_6.1}. \textit{(Right)} Incoherent scattering functions $S(q,Z)$ for the same elements (dots), and $|Z - F(Z,q)|^2$ factors (lines). The incoherent scattering functions are taken from Ref.~\cite{10.1063/1.555523}\label{fig:Fq}. The comparison indicates the momentum-transfer region where inelastic (incoherent) scattering, neglected in Eq.~\ref{eq:invP_dsigma}, becomes relevant (see text).}\label{fig:formfac}
\end{figure*}

In this work, we model the atomic form factor $F(q)$ using Relativistic Hartree--Fock (RHF) calculations, which provide an accurate description of electron screening at keV-scale momentum transfers. Following Ref.~\cite{ITC_VolC_6.1}, we parametrize the form factor as
\begin{equation}
    F(q) = \sum_{i} a_i \exp\left[-b_i \left(\frac{q}{4\pi}\right)^2 \right] + c \, ,
    \label{eq:RHF}
\end{equation}
where the coefficients $(a_i, b_i, c)$ are specific to each atomic species (Pb, W, O). We show the resulting atomic form factors in the left panel of Fig.~\ref{fig:formfac}.

The total cross section is obtained by integrating Eq.~(\ref{eq:invP_dsigma}) over the full solid angle,
\begin{equation}
    \sigma_{P}(E_a) = 2\pi \int_{0}^{\pi} d\theta \,
    \sin\theta \, \frac{d\sigma_{P}}{d\Omega} \, .
    \label{eq:sigma_tot}
\end{equation}

\begin{figure}[]
\centering
\includegraphics[width=.89\textwidth]{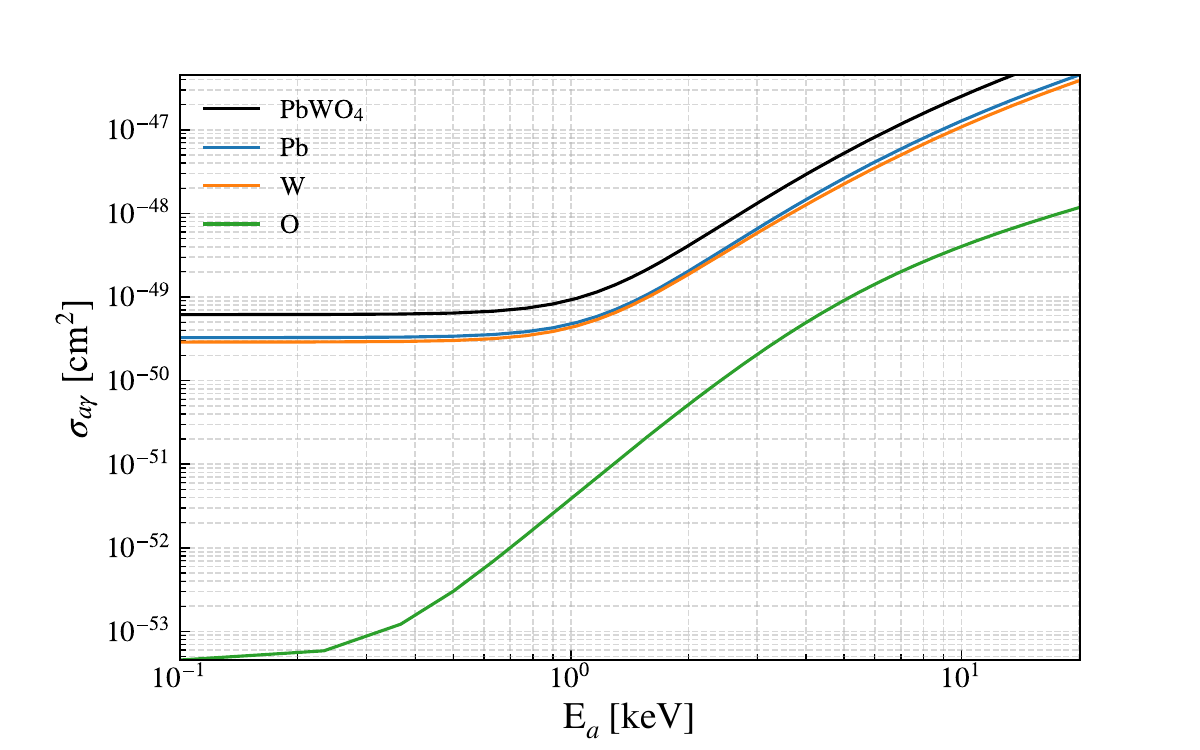}
\caption{Inverse Primakoff cross-section in Pb, W and O assuming $g_{a\gamma} = 10^{-10}$\,GeV$^{-1}$. The cross-section for PbWO$_4$ was obtained as the sum of the elemental contributions weighted by their stoichiometric coefficients.\label{fig:invP}}
\end{figure}

The integral is dominated by small scattering angles, reflecting the forward-peaked nature of the process, with the physical forward divergence being regulated by the atomic form factor.

The cross section in Eq.~\eqref{eq:invP_dsigma} accounts only for elastic (coherent) scattering,
in which the atom is left in its ground state. Following Ref.~\cite{Abe:2020mcs} , the relevance
of inelastic channels, in which the final state contains an excited or ionized
atom, can be assessed by comparing the incoherent scattering function
$S(q,Z)$~\cite{10.1063/1.555523} with the coherent factor $|Z-F(q)|^{2}$, as shown in the right
panel of Fig.~\ref{fig:formfac}. For momentum transfers $q \gtrsim 1$~keV, which dominate the
cross-section integral for the axion energies relevant to this search, the
coherent term prevails and the elastic treatment is accurate. At lower momentum
transfers the incoherent contribution becomes comparable or even dominant~\cite{Abe:2020mcs};
neglecting it only underestimates the expected rate, making our results
conservative. This region is relevant only for the sub-keV portion of the
spectrum, most notably the LP component, which does not contribute to the
exclusion reach derived below (see Sec.~\ref{sec:results}) and is retained as a discovery
target for a future low-threshold configuration.

The resulting cross sections for Pb, W, and O are shown in Fig.~\ref{fig:invP}, assuming $g_{a\gamma} = 10^{-10}\, \mathrm{GeV}^{-1}$. As expected, the contribution is strongly enhanced for high-$Z$ nuclei, making Pb and W the dominant targets in PbWO$_4$.

The differential interaction rate per unit detector mass is computed by folding the cross section with the incoming axion flux:
\begin{equation}
    \frac{dR_P}{dE_a} = \frac{N_A}{M_{\rm mol}} \, \frac{ d\Phi(E_a)}{dE_a} \sum_{i=\mathrm{Pb,W,O}}
      w_i\,\sigma_{P,i}(E_a) \, ,
    \label{eq:rate}
\end{equation}
where $N_A$ is Avogadro's number, $M_{\rm mol}$ is the molar mass of PbWO$_4$, and $w_i$ are the stoichiometric weights in PbWO$_4$. The rate is expressed in units of events per keV per kg per day.

Finally, we remark that the inverse Primakoff process is directly related to Primakoff production ($\gamma \to a$), with the two cross sections differing only by a factor of two due to photon polarization averaging. This relation provides a useful consistency check when comparing production and detection calculations.

\subsection{Axioelectric Effect}

Axions interacting with electrons can be detected via the axioelectric effect. This corresponds to the absorption of an axion by a bound electron in an atom with atomic mass number $A$, resulting in the ejection of the electron, in close analogy with the photoelectric effect,
\begin{equation}
    a + A \rightarrow e^- + A^+ \, .
\end{equation}
This process provides one of the most efficient detection channels for axions with energies in the keV range, particularly in materials with large atomic numbers.

In the non-relativistic limit $E_a\ll m_e$, the axioelectric cross section is related to the photoelectric cross section $\sigma_{pe}$ through~\cite{Pospelov:2008jk,Derevianko:2010kz}
\begin{equation}
    \sigma_{ae}(E_a) = \sigma_{pe}(E_a) \,
    \frac{g_{ae}^2}{\beta_a} \,
    \frac{3 E_a^2}{16\pi \alpha m_e^2}
    \left(1 - \frac{\beta_a^{2/3}}{3} \right) \, ,
    \label{eq:sigmaAE_full}
\end{equation}
where $\alpha$ is the fine-structure constant, $m_e$ is the electron mass, and $\beta_a = v_a/c$ is the axion velocity.

\begin{figure}[]
\centering
\includegraphics[width=.89\textwidth]{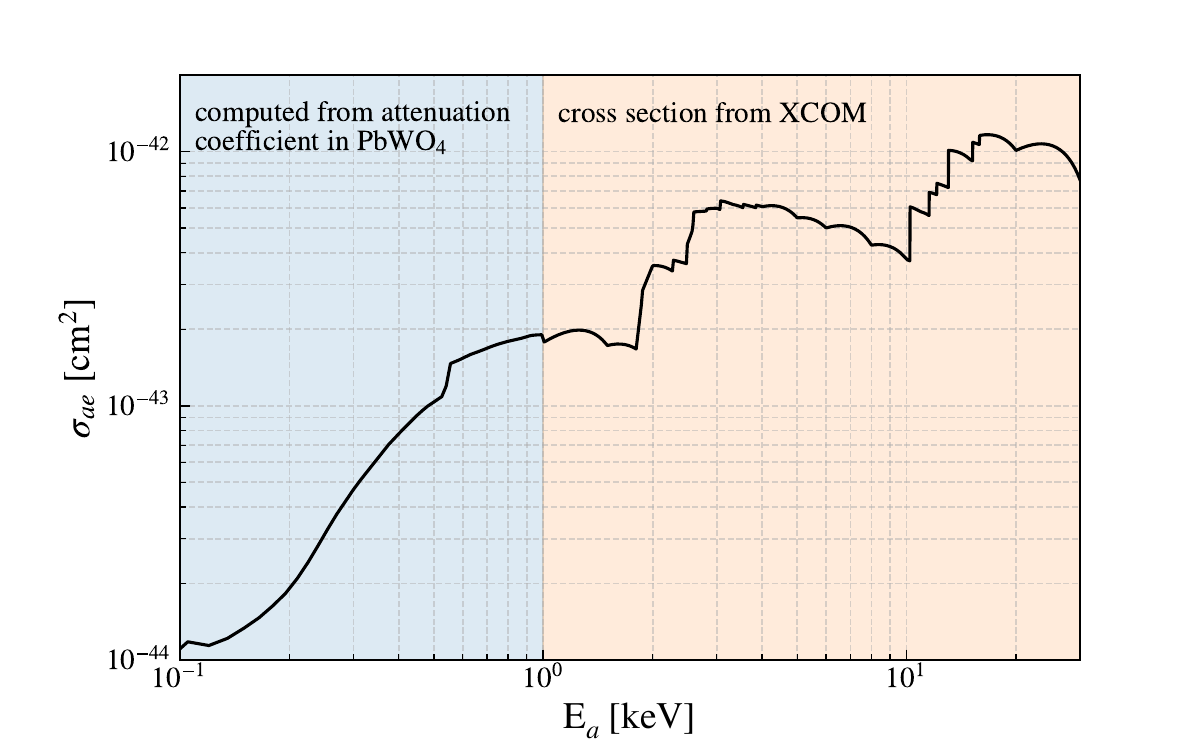}
\caption{Axioelectric cross section in PbWO$_4$ assuming g$_{ae} = 5 \times 10^{-11}$. The cross-section was calculated adapting the photoelectric cross-section, see Eq.~\eqref{eq:sigmaAE_full}. The latter was taken from Ref.~\cite{NIST_XCOM} for energies greater than 1~keV, while for $E_a\lesssim1\,{\rm keV}$ it was determined starting from the X-ray mass attenuation coefficients for Pb, W and O from Ref.~\cite{NIST_Hubbell}.\label{fig:sigmaAE}}
\end{figure}

This relation highlights that the axioelectric effect inherits the strong atomic dependence of the photoelectric effect, which scales approximately as $\sigma_{pe} \propto Z^n$ with $n \sim 4$--$5$ in the keV regime. As a consequence, high-$Z$ materials such as Pb and W provide a substantial enhancement of the interaction probability, making PbWO$_4$ an excellent target for this channel.

In practice, the photoelectric cross section is obtained from tabulated atomic data. In this work, we use values from the NIST XCOM database~\cite{NIST_XCOM} for energies above 1~keV. At lower energies, the cross section is reconstructed from X-ray mass attenuation coefficients~\cite{NIST_Hubbell}. For a compound material such as PbWO$_4$, the total cross section is computed as a weighted sum of the elemental contributions,
\begin{equation}
    \sigma_{pe}^{\mathrm{PbWO}_4}(E_a) =
    \sum_{i=\mathrm{Pb,W,O}} w^{'}_i \, \sigma_{pe}^{(i)}(E_a) \, ,
\end{equation}
where $w^{'}_i$ are the mass fractions of each element.

As for the Primakoff detection rate, the differential interaction rate per unit detector mass is then obtained by folding the cross section with the incoming axion flux,
\begin{equation}
    \frac{dR_{ae}}{dE_a} =
    \frac{N_A}{M_{\rm mol}} \,
    \frac{d\Phi(E_a)}{dE_a}  \,
    \sigma_{ae}(E_a) \, ,
\end{equation}
where $M_{mol}$ is the molar mass of the target.

A distinctive feature of the axioelectric effect is that, unlike scattering processes, the full axion energy is transferred to the electron (up to the binding energy), resulting in a mono-energetic deposition for line-like axion sources and a spectrum that directly traces the incoming axion flux for continuous sources. This property, combined with the steep energy dependence of $\sigma_{ae}$, makes this channel particularly powerful for probing solar axions in the keV range.

The resulting axioelectric cross section for PbWO$_4$ is shown in Fig.~\ref{fig:sigmaAE}, assuming $g_{ae} = 5 \times 10^{-11}$. As expected, the cross section increases rapidly with energy and is dominated by the contribution of the heaviest elements, reflecting the strong $Z$ dependence of the underlying photoelectric process.

\section{Expected Signal and Detector Response}\label{sec:detres}
The expected interaction rate in RES-NOVA is determined by the convolution of the production fluxes and the detection cross-sections. A key feature of this search is the \textit{cross-role} of the coupling constants: $g_{a\gamma}$ and $g_{ae}$ govern both the production in the solar core and the detection in the PbWO$_4$ crystals, while $g_{aN}^{\rm eff}$ influences only the axion production.

The total differential interaction rate $dR/dE$ (counts/keV/kg/s) can be expressed as: 
\begin{equation}
    \frac{dR}{dE_a} = \frac{N_A}{M_{\rm mol}} \frac{d\Phi(E_a)}{dE_a}  \times \sigma(E_a)\,,
\end{equation}
where $d\Phi/dE_a$ and $\sigma$ represent the total solar axion flux and the total cross-section, whose contributions depend on the non-vanishing axion couplings assumed in the analysis. Specifically, the interaction rate depends on a non-trivial combination $\propto g_{ai}^2\,g_{aj}^2$, where $g_{ai}=\{g_{a\gamma},\,g_{ae},\,g_{aN}^{\rm eff}\}$ are the coupling relevant for production and $g_{aj}=\{g_{a\gamma},\,g_{ae}\}$ are relevant for detection.

To account for the experimental response, the theoretical rate is convolved with the detector's energy resolution $\sigma_E$ and multiplied by the detection efficiency $\mathcal{E}(E)$: 
\begin{equation}
    \frac{dS(E)}{dE} = \mathcal{E}(E)\label{eq:detres}
    \int \frac{dR}{dE_a} \frac{1}{ \sqrt{2\pi}\sigma_E } e^{-\frac{1}{2}\left(\frac{E-E_a}{\sigma_E}\right)^2}dE_a\,.
\end{equation}
In this study, we consider two benchmark energy resolutions, $\sigma_E = 0.2$~keV and $\sigma_E = 0.02$~keV, to evaluate the detector's performance in resolving solar axion signatures from the background. In cryogenic detectors of this type the analysis energy threshold is conventionally set a few baseline-noise standard deviations above zero; adopting $E_{\mathrm{thr}} \simeq 5\,\sigma_E$, the two benchmark resolutions correspond to thresholds of $\simeq 1$~keV and $\simeq 100$~eV, respectively. We account for the energy detection threshold including it in the efficiency $\mathcal{E}(E)=\Theta(E-E_{\rm thr})$. The total number of expected signal counts in an energy bin $i$ is then 
\begin{equation}
S_i = M\,T\int_{\Delta E_i} \frac{dS(E)}{dE}\, dE  \,, \label{eq:Si}
\end{equation}
where $M$ is the mass and $T$ the exposure time. This detailed modeling of the spectral shape is essential for likelihood analysis in order to disentangle the contributions of the different couplings.

\section{Sensitivity projection methodology}
\label{sec:projection}

The sensitivity of the RES-NOVA demonstrator is evaluated with a frequentist
profile-likelihood-ratio test on an \textit{Asimov dataset} $\mathcal{D}_A$, in
which the observed counts in each energy bin equal the background expectation,
$n_i = B_i$, with $B_i$ taken from the Monte Carlo background
model~\cite{RES-NOVACollaboration:2025stq}.

The expected signal enters through the counts $S_i$ of Eq.~\eqref{eq:Si}, which
depend on the full set of axion couplings probed in this work,
\begin{equation}
    \vec{g} \equiv (g_{ae},\, g_{a\gamma},\, g_{aN}^{\rm eff}) \, ,
\end{equation}
where $g_{ae}$ and $g_{a\gamma}$ govern the continuous Primakoff, LP and ABC
components through the axioelectric and inverse Primakoff channels
(Sec.~\ref{sec:detres}), while $g_{aN}^{\rm eff}$ sets the intensity of the $^{57}$Fe nuclear
line at $14.4$~keV. The binned Poisson likelihood then reads
\begin{equation}
    \mathcal{L}(\vec{g}\,|\,\mathcal{D}_A)
    =
    \prod_{i}
    \frac{\big(S_i(\vec{g}) + B_i\big)^{B_i}}{B_i!}\,
    e^{-(S_i(\vec{g}) + B_i)} \, ,
    \label{eq:likelihood}
\end{equation}
and the exclusion limits follow from the likelihood-ratio test statistic
\begin{equation}
    \Lambda(\vec{g})
    =
    -2 \ln
    \frac{\mathcal{L}(\vec{g}\,|\,\mathcal{D}_A)}
         {\mathcal{L}(\hat{\vec{g}}\,|\,\mathcal{D}_A)} \, ,
    \label{eq:teststat}
\end{equation}
where $\hat{\vec{g}}$ is the best-fit coupling set. For the background-only Asimov
dataset the maximum lies at the null-signal hypothesis, $\hat{\vec{g}} = 0$, since
any non-zero coupling adds signal and lowers the likelihood.

Limits are derived in the three two-dimensional coupling planes,
$(g_{ae},\, g_{a\gamma})$, $(g_{ae},\, g_{aN}^{\rm eff})$ and
$(g_{a\gamma},\, g_{aN}^{\rm eff})$. In each plane the two couplings of interest are
scanned while the third is held fixed at zero: $g_{aN}^{\rm eff}=0$ in the
$(g_{ae},\, g_{a\gamma})$ plane, $g_{a\gamma}=0$ in the
$(g_{ae},\, g_{aN}^{\rm eff})$ plane, and $g_{ae}=0$ in the $(g_{a\gamma},\, g_{aN}^{\rm eff})$
plane. The 90\% C.L.\ regions are defined by $\Lambda \leq 4.61$, as expected for a
$\chi^2$-distributed test statistic with two degrees of freedom.

\section{Exclusion limit from prototype data}
\label{sec:prototype}

The projections of Sec.~\ref{sec:projection} are complemented in this work by a
constraint derived from real data: an exclusion limit in the
$(g_{ae},\,g_{a\gamma})$ plane, shown as the green contour in
Fig.~\ref{fig:sensitivity}. The dataset was collected with a $13\,\mathrm{g}$
PbWO$_4$ crystal grown from archaeological Pb and operated as a cryogenic
calorimeter at the underground Gran Sasso Laboratory of INFN (Italy)---the same measurement
recently used to set the first dark-matter exclusion limits and investigation of inelastic dark-matter interactions with this
compound~\cite{RES-NOVA:2026prototype, Alloni:2026xdf}. The detector, read out with a Ge-NTD
thermistor, accumulated an exposure of $32.4\,\mathrm{g\cdot day}$ with an
analysis threshold of $2.5$~keV and a baseline energy resolution of
$\sigma = 234$~eV; a complete description of the experimental setup, data
processing and detector performance is given in
Ref.~\cite{RES-NOVA:2026prototype}.

Given the prototype and R\&D nature of this measurement, a detailed and reliable
background model is not available, so the likelihood approach of
Sec.~\ref{sec:projection} cannot be applied to these data. To obtain robust and
conservative limits under these conditions we employ the optimum-interval method,
commonly referred to as the Yellin method~\cite{Yellin:2002xd}. This approach
allows one to derive exclusion limits without relying on an explicit background
prediction, and is therefore particularly well suited for datasets lacking precise modeling.
While routinely used in direct dark matter searches, its application to a
multi-coupling axion parameter space is novel.

We evaluate the exclusion directly on a two-dimensional grid spanning the
$(g_{ae},g_{a\gamma})$ plane, treating the two couplings jointly. At each grid
node, i.e.\ for each pair $(g_{ae},g_{a\gamma})$, the expected signal $S(E)$ of
Eq.~\eqref{eq:detres} provides the two ingredients of the analysis. The first is the
normalized cumulative distribution function (CDF)
\begin{equation}
    x(E;\,g_{ae},g_{a\gamma})
    =
    \frac{\int_{E_{min}}^{E} \frac{dS}{dE'}(E';\,g_{ae},g_{a\gamma})\,dE'}
         {\int_{E_{min}}^{E_{max}} \frac{dS}{dE'}(E';\,g_{ae},g_{a\gamma})\,dE'} \, ,
    \label{eq:cdf}
\end{equation}
which transforms the data into a uniform variable $x\in[0,1]$, as required by the
optimum-interval construction. Each observed event is mapped onto its value of $x$
through this CDF, and the Yellin algorithm is applied to the transformed data,
returning the upper bound $N_{\mathrm{sup}}(g_{ae},g_{a\gamma})$ on the number
of signal events compatible with the observed distribution at the chosen confidence
level (C.L.), here $90\%$. The second is the total expected number of signal events,
\begin{equation}
    N(g_{ae},g_{a\gamma})
    =
    \sum_i S_i
    =
    M\,T\int_{E_{min}}^{E_{max}} \frac{dS}{dE}(E;\,g_{ae},g_{a\gamma})\,  \, dE \, ,
    \label{eq:Npred}
\end{equation}
i.e.\ the sum over bins of the counts $S_i$ already defined in
Sec.~\ref{sec:detres}. 

Repeating this operation over the grid produces the matrix
$N_{\mathrm{sup}}(g_{ae},g_{a\gamma})$, which is then compared node by node
with the predicted counts $N(g_{ae},g_{a\gamma})$ of Eq.~\eqref{eq:Npred}. A
given pair of couplings is excluded at $90\%$ C.L.\ when the predicted signal
exceeds the optimum-interval bound,
\begin{equation}
    N(g_{ae},g_{a\gamma})
    \;\ge\;
    N_{\mathrm{sup}}(g_{ae},g_{a\gamma}) \, ,
    \label{eq:exclusion}
\end{equation}
so that thresholding the matrix at the equality $N = N_{\mathrm{sup}}$ directly
traces the $90\%$ C.L.\ exclusion contour in the $(g_{ae},g_{a\gamma})$ plane.

The two ingredients play complementary roles. By construction the CDF is insensitive
to the overall normalization of the signal, and therefore to the global coupling
scale; it depends only on the spectral shape, which is fixed by the relative weight
of the $g_{ae}^4$, $g_{ae}^2 g_{a\gamma}^2$ and $g_{a\gamma}^4$ terms
introduced in Sec.~\ref{sec:detres}, i.e.\ by the coupling ratio $g_{a\gamma}/g_{ae}$. Consequently the
Yellin bound $N_{\mathrm{sup}}$ is constant along rays of fixed coupling ratio
emanating from the origin, whereas the predicted counts $N$ grow monotonically along
each such ray. Every ray therefore crosses the condition $N = N_{\mathrm{sup}}$ at a
single point, and the locus of these crossings defines the exclusion boundary. In
this way the spectral information enters through $N_{\mathrm{sup}}$ while the
absolute rate enters through $N$, and the two are consistently combined at every
point of the parameter space.
 
\section{Results}
\label{sec:results}

The two main results of our analysis are the \emph{projected} sensitivity of the RES-NOVA demonstrator and the \emph{measured} exclusion limit obtained from the prototype data. The former is derived with the profile-likelihood analysis described in Sec.~\ref{sec:projection}, assuming an exposure of $1\,\mathrm{ton\cdot y}$, corresponding to the $170$~kg demonstrator operated for $\simeq 6$~years. The latter is obtained from the $13\,\mathrm{g}$ prototype data and sets an exclusion limit in the $(g_{ae},g_{a\gamma})$ plane using the optimum-interval method described in Sec.~\ref{sec:prototype}.

\subsection{Projected sensitivity}
\label{sec:results_projection}

\begin{figure}[t!]
\centering
\includegraphics[width=.89\textwidth]{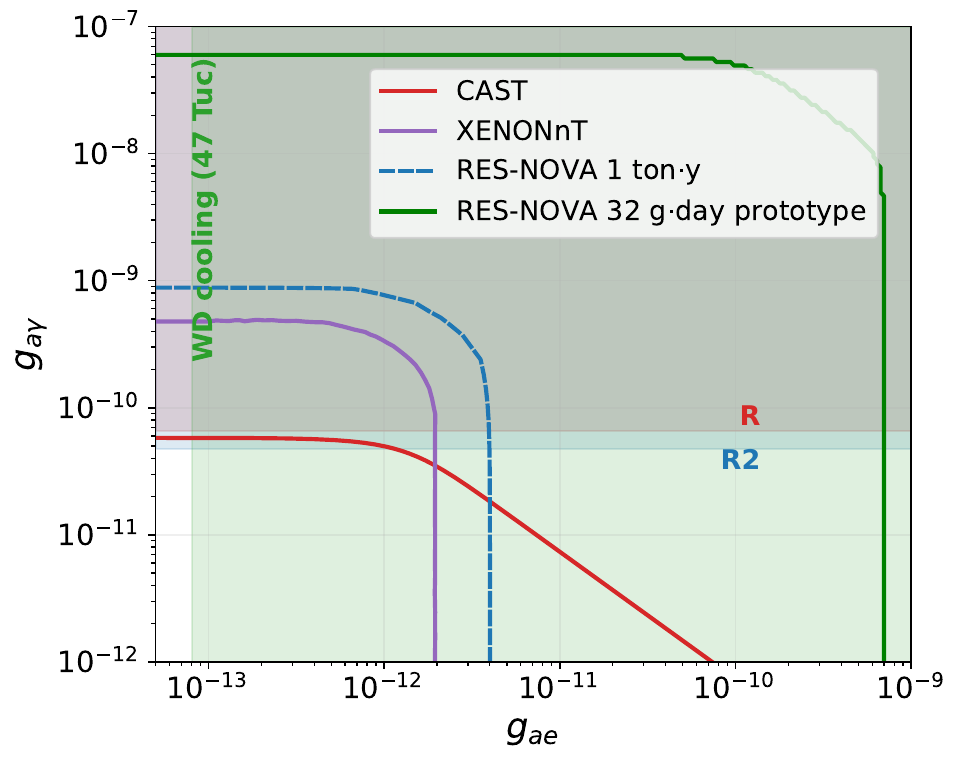}
\caption{Projected RES-NOVA sensitivity and measured prototype exclusion in the
$(g_{ae},\,g_{a\gamma})$ plane at $90\%$ C.L.; for each curve the region towards
larger couplings is excluded. The blue curve is the \emph{projected} reach for a
$1\,\mathrm{ton\cdot y}$ exposure, from the profile-likelihood analysis on the Asimov
dataset (Sec.~\ref{sec:projection}). The green curve is the exclusion limit
\emph{measured} with the $13\,\mathrm{g}$ PbWO$_4$ prototype
($32.4\,\mathrm{g\cdot day}$ exposure)~\cite{RES-NOVA:2026prototype}, set with the
optimum-interval method (Sec.~\ref{sec:prototype}). Existing constraints from the CAST
helioscope~\cite{CAST:2024} and XENONnT~\cite{XENON:2022ltv} are shown for
comparison. We display also astrophysical bounds on $g_{a\gamma}$ from the $R$~\cite{HBstars_Ayala2014} and $R_2$~\cite{R2_Dolan2022} parameters in globular clusters R2 and on $g_{ae}$ from white dwarf cooling in 47 Tucanae~\cite{WD47Tuc_Fleury2025}.\label{fig:sensitivity}}
\end{figure}

Figure~\ref{fig:sensitivity} shows the projected RES-NOVA sensitivity (dashed blue line) and the current limit from the prototype data (solid green line) in the $(g_{ae},g_{a\gamma})$ plane. For comparison, we also show astrophysical limits on $g_{ae}$ from white-dwarf cooling in 47 Tucanae~\cite{WD47Tuc_Fleury2025} and on $g_{a\gamma}$ from the $R$~\cite{HBstars_Ayala2014} and $R_2$~\cite{R2_Dolan2022} parameters in globular clusters, as well as existing experimental constraints~\cite{XENON:2022ltv,CAST:2024}.

The shape of the projected sensitivity and of the current exclusion contour in the $(g_{ae},g_{a\gamma})$ plane is determined by the interplay between axion production and detection, since both couplings can contribute to the solar axion flux and to the detection rate. For $g_{ae}\lesssim 10^{-12}$, the RES-NOVA demonstrator is expected to be sensitive to $g_{a\gamma}\gtrsim 10^{-9}\,\mathrm{GeV}^{-1}$, essentially independently of $g_{ae}$, since production and detection channels controlled by the axion--electron coupling become inefficient. Conversely, for $g_{a\gamma}\lesssim 10^{-10}\,\mathrm{GeV}^{-1}$, values of $g_{ae}\gtrsim 4\times 10^{-12}$ can be probed.

The sensitivity in this plane is mainly driven by the ABC and Primakoff fluxes, which peak at $E_a\simeq 1$~keV and $E_a\simeq 4.17$~keV, respectively, and therefore deposit most of their spectral weight above $1$~keV. Improving the energy resolution from $\sigma_E=0.2$\,keV to $\sigma_E=0.02$\,keV, and correspondingly lowering the threshold from $\simeq 1$~keV to $\simeq 100$~eV (see Sec.~\ref{sec:detres}), leaves the exclusion contours nearly unchanged. This confirms that a sub-keV threshold does not significantly enhance the sensitivity in the $(g_{ae},g_{a\gamma})$ plane. The LP component, which is confined to $E_a\lesssim 200$~eV, does not contribute appreciably to the exclusion reach. Nevertheless, it represents an important discovery target: while a $1$~keV threshold is sufficient to set competitive limits, resolving LP axions in the event of a positive signal, and thereby accessing information on the magnetic-field structure of the deep solar interior, would require the sub-$200$~eV threshold enabled by the finer-resolution configuration.

\begin{figure*}[t!]
\centering
\includegraphics[width=.44\textwidth]{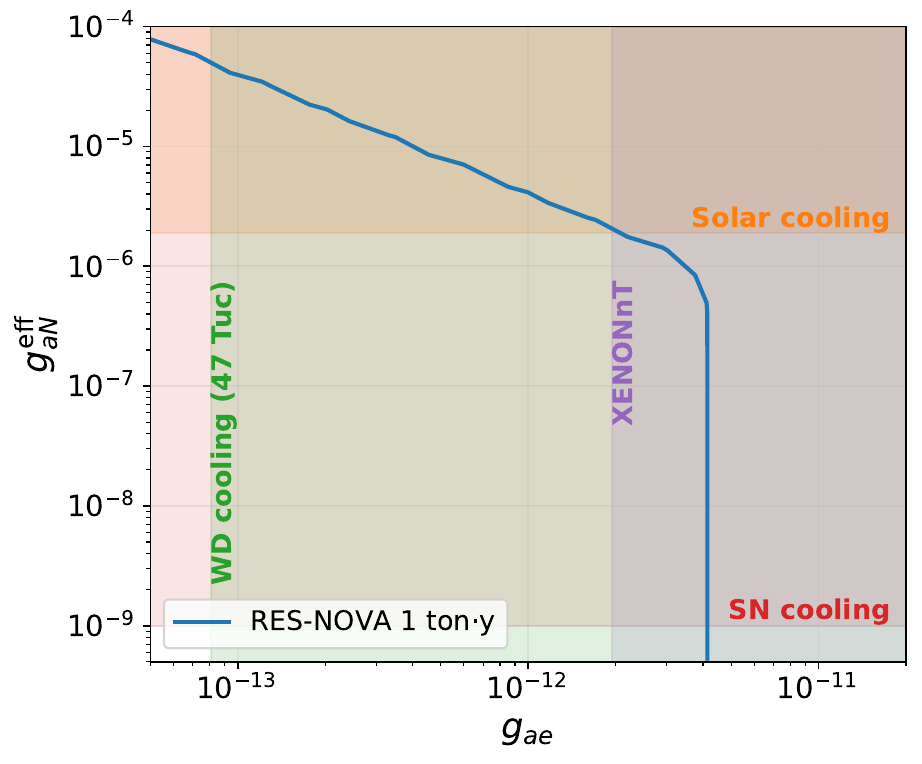} \;\hfill
\includegraphics[width=.44\textwidth]{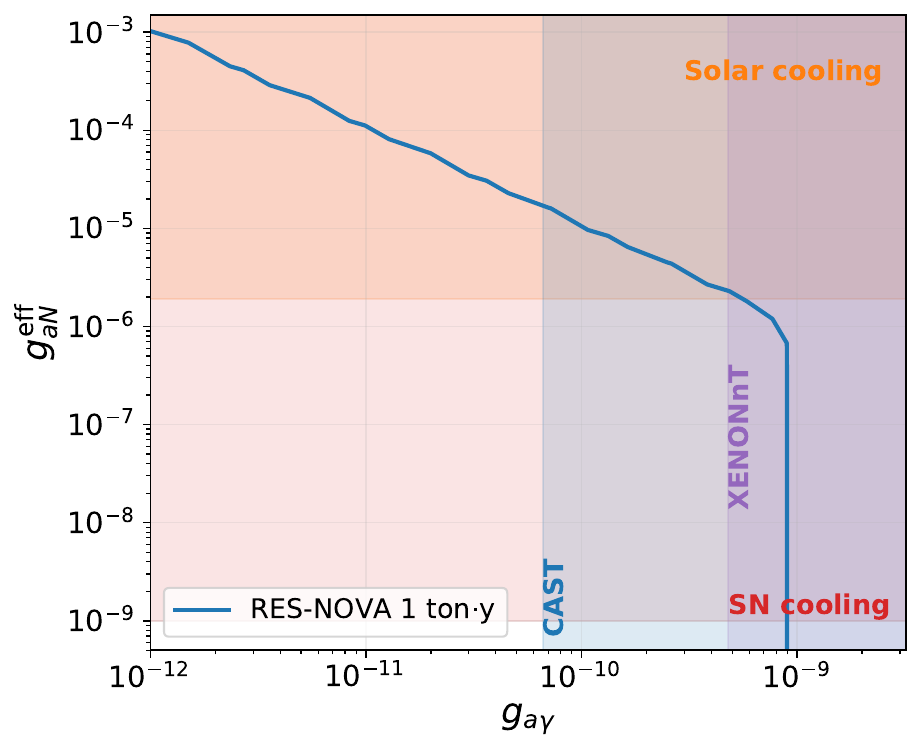}
\caption{Projected RES-NOVA sensitivity ($1\,\mathrm{ton\cdot y}$ exposure, $90\%$
C.L.) to the axion--nucleon coupling, probed through the $^{57}$Fe $14.4$~keV line.
\textit{(Left)} $(g_{ae},\,g_{aN}^{\rm eff})$ plane, with $g_{a\gamma}=0$.
\textit{(Right)} $(g_{a\gamma},\,g_{aN}^{\rm eff})$ plane, with $g_{ae}=0$. In each panel
the region towards larger couplings is excluded. Shaded regions indicate parameter
space already excluded by existing astrophysical constraints. The horizontal bands
correspond to the solar-cooling bound $g_{aN}^{\rm eff}\le1.89\times10^{-6}$~\cite{gaN_DiLuzio2022}
and the SN~1987A bound $g_{aN}^{\rm eff}\lesssim1.0\times10^{-9}$~\cite{Lella:2022uwi}.
The vertical shaded band in the right panel marks the CAST bound on
$g_{a\gamma}$~\cite{CAST:2024}, while that in the left panel marks the
47~Tucanae white-dwarf cooling bound on $g_{ae}$~\cite{WD47Tuc_Fleury2025}.}\label{fig:gaN}
\end{figure*}

In Fig.~\ref{fig:gaN}, we show the projected RES-NOVA sensitivity in the $(g_{ae},g_{aN}^{\rm eff})$ plane (left panel) and in the $(g_{a\gamma},g_{aN}^{\rm eff})$ plane (right panel). We compare these projections with existing experimental limits~\cite{XENON:2019gfn,CAST:2024} and astrophysical bounds on $g_{ae}$~\cite{WD47Tuc_Fleury2025}, $g_{a\gamma}$~\cite{HBstars_Ayala2014,R2_Dolan2022}, and $g_{aN}^{\rm eff}$ from supernovae~\cite{Lella:2022uwi,Lella:2023bfb}\footnote{A bound comparable to the supernova cooling constraint is obtained also using neutron stars~\cite{Buschmann:2021juv}.} and solar cooling~\cite{gaN_DiLuzio2022}. The axion--nucleon coupling $g_{aN}^{\rm eff}$ is probed exclusively through production of the monochromatic $^{57}$Fe line at $14.4$~keV. Therefore, RES-NOVA cannot constrain $g_{aN}^{\rm eff}$ alone: sensitivity in these planes requires a nonzero axion--electron or axion--photon coupling to provide an efficient absorption channel in the detector.

\subsection{Prototype exclusion limit}
\label{sec:results_prototype}

The current prototype data already yield a genuine experimental constraint in the $(g_{ae},g_{a\gamma})$ plane. In particular, the prototype excludes $g_{ae}\gtrsim 7\times 10^{-10}$ for $g_{a\gamma}\lesssim 10^{-8}\, \mathrm{GeV}^{-1}$, and $g_{a\gamma}\gtrsim 6\times 10^{-8}\, \mathrm{GeV}^{-1}$ for $g_{ae}\lesssim 10^{-10}$ (see the green contour in Fig.~\ref{fig:sensitivity}). Despite the modest exposure and the absence of a background model, this represents the first solar-axion limit obtained on real data with a cryogenic PbWO$_4$ crystal grown from archaeological Pb~\cite{RES-NOVA:2026prototype}. The prototype limit is not shown in the planes of Fig.~\ref{fig:gaN}, since its $32.4\,\mathrm{g\cdot day}$ exposure gives negligible sensitivity to the $^{57}$Fe line.

\subsection{Comparison with existing experimental constraints}
\label{sec:results_comparison}

It is useful to compare the projected RES-NOVA sensitivity with existing experimental limits from XENONnT~\cite{XENON:2022ltv} and CAST~\cite{CAST:2024}. Although the background level achieved by XENONnT is roughly four orders of magnitude lower, the projected RES-NOVA sensitivity in the $(g_{ae},g_{a\gamma})$ plane is weaker only by a factor of $\sim 2$. This is mainly due to the large target mass of the RES-NOVA demonstrator, and the high-$Z$ enhancement of the axioelectric and inverse-Primakoff cross sections in Pb and W. The CAST helioscope sets a stronger absolute limit on $g_{a\gamma}$, but this bound relies on coherent $a\to\gamma$ conversion and is therefore valid only for $m_a\lesssim 0.02$~eV in vacuum, degrading at higher masses except in narrow buffer-gas-tuned windows~\cite{CAST:2024}. By contrast, the RES-NOVA constraint is based on inverse-Primakoff absorption and does not require a conversion baseline. It can therefore be extended to axion masses up to $m_a\lesssim T_c\sim 1$~keV.

\section{Conclusions}
\label{sec:conclusions}

We have carried out a detailed study of the sensitivity of RES-NOVA to solar axions in
the keV energy range. Starting from the solar production fluxes---the Primakoff and
longitudinal-plasmon (LP) components governed by $g_{a\gamma}$, the ABC flux
governed by $g_{ae}$, and the $^{57}$Fe nuclear line governed by $g_{aN}^{\rm eff}$---we computed
the expected signal in PbWO$_4$ through the two dominant detection channels, inverse
Primakoff conversion and the axioelectric effect, folding the resulting rate with the
detector energy resolution and efficiency. The interplay of production and detection, in
which $g_{ae}$ and $g_{a\gamma}$ enter at both stages, produces the characteristic
$g_{a\gamma}^4$,\,$g_{ae}^4$ and $g_{a\gamma}^2 g_{ae}^2$,\, ${g_{aN}^{\rm eff}}^2 g_{ae}^2$,\, ${g_{aN}^{\rm eff}}^2 g_{a\gamma}^2$ dependence that lets a single experiment probe a broad region of the
multi-coupling parameter space.

On this basis, this work delivers two complementary results. The first is a
\emph{projection}: the experimental reach of the RES-NOVA demonstrator, yielding
$90\%$ C.L.\ exclusions for a $1\,\mathrm{ton\cdot y}$ exposure in the
$(g_{ae},\,g_{a\gamma})$, $(g_{ae},\,g_{aN}^{\rm eff})$ and
$(g_{a\gamma},\,g_{aN}^{\rm eff})$ planes. The second is a \emph{measured
constraint}: the first solar-axion exclusion limit obtained on real data with a
cryogenic PbWO$_4$ detector grown from archaeological
lead~\cite{RES-NOVA:2026prototype}, establishing the technology as a viable avenue
for these searches. As discussed in Sec.~\ref{sec:results}, the projected reach in
$g_{ae}$ comes within a factor of $\sim 2$ of the current best direct bound despite
a substantially higher background and lower exposure, while the mass-independent bound on $g_{a\gamma}$ is complementary to the
stronger but mass-limited constraints of magnetic helioscopes such as CAST and the
upcoming IAXO~\cite{CAST:2024,IAXO:2025ltd}. Moreover, the simultaneous sensitivity
to $g_{ae}$, $g_{a\gamma}$ and $g_{aN}^{\rm eff}$ is the prerequisite for eventually
testing the coupling pattern that distinguishes hadronic (KSVZ-like) from
non-hadronic (DFSZ-like) models.

Looking ahead, several developments will extend this reach. The demonstrator is a
step toward the full $1.8\,\mathrm{ton}$ RES-NOVA experiment~\cite{Pattavina:2020cqc},
whose roughly tenfold larger mass, and the correspondingly larger exposure, will improve
the statistical reach in all three coupling planes. Because this projection is
background-limited, the most effective lever is background reduction: further
improvements in crystal radiopurity would translate
directly into stronger exclusions. Together, larger exposure and lower background would
push the projected reach toward the parameter space favored by benchmark axion models. A
dedicated low-threshold configuration, exploiting the sub-$200$~eV thresholds already
within reach of the TES technology~\cite{FerreiroIachellini:2021qgu}, would additionally
open the longitudinal-plasmon window discussed in Sec.~\ref{sec:results} --- and with it a handle on the deep solar magnetic field. Taken together, these results demonstrate the potential of RES-NOVA as a
competitive, multi-coupling solar axion experiment.

\acknowledgments

N.~Ferreiro Iachellini acknowledges support from the COST Action Cosmic WISPers (CA21106), through Short Term Scientific Mission grant E-COST-GRANT-CA21106-04f918d0, which funded a research stay at the University of Zaragoza, and thanks the University of Zaragoza for its hospitality during this visit. He also thanks Sophia Hollick for sharing her expertise regarding axion interactions in direct detection experiments and Cristina Margalejo Blasco for her helpful insights in CAST data and its interpretation. The RES-NOVA experiment is funded by the European Research Council (ERC) under the European Union's Horizon Europe research and innovation programme, Grant Agreement No.~101087295 (RES-NOVA). We gratefully acknowledge the University of Milano--Bicocca and INFN for their continuous support of the collaboration. S.~Ghislandi acknowledges support from the U.S. Department of Energy (DOE) Grant No. [DE-SC0011091]. 
MG acknowledges support from the Spanish Agencia Estatal de Investigación under grant PID2019-108122GB-C31, funded by MCIN/AEI/10.13039/501100011033, and from the “European Union NextGenerationEU/PRTR” (Planes complementarios, Programa de Astrofísica y Física de Altas Energías). He also acknowledges support from grant PGC2022-126078NB-C21, “Aún más allá de los modelos estándar,” funded by MCIN/AEI/10.13039/501100011033 and “ERDF A way of making Europe.” Additionally, MG acknowledges funding from the European Union’s Horizon 2020 research and innovation programme under the European Research Council (ERC) grant agreement ERC-2017-AdG788781 (IAXO+). 
MG and VF acknowledge support from project CNS2025-165965, “Señales de axiones en el rango MeV desde fuentes astrofísicas hasta detectores actuales y futuros”, funded by MICIU/AEI/10.13039/501100011033. VF acknowledges support from the Universidad de Zaragoza under the predoctoral contract call PI-PRD/2024-001.
AL are supported by the Italian MUR through the FIS 2 project FIS-2023-01577 (DD n. 23314 10-12-2024, CUP C53C24001460001), and by Istituto Nazionale di Fisica Nucleare (INFN) through the Theoretical Astroparticle Physics (TAsP) project. G.L. acknowledges support from the U.S. Department of Energy under contract number DE-AC02-76SF00515.

\bibliographystyle{JHEP}
\bibliography{main}

\end{document}